\documentclass[twocolumn,preprintnumbers,amsmath,amssymb]{revtex4}

\usepackage{graphicx}
\usepackage{dcolumn}
\usepackage{bm}

\newcommand{\nc}{\newcommand}

\nc{\ra}{\rightarrow}

\nc{\da}{\dagger}
\nc{\dn}{\downarrow}
\nc{\cA}{{\cal A}}
\nc{\cB}{{\cal B}}
\nc{\cC}{{\cal C}}
\nc{\cD}{{\cal D}}
\nc{\cE}{{\cal E}}
\nc{\cF}{{\cal F}}
\nc{\cG}{{\cal G}}
\nc{\cH}{{\cal H}}
\nc{\cI}{{\cal I}}
\nc{\cJ}{{\cal J}}
\nc{\cK}{{\cal K}}
\nc{\cL}{{\cal L}}
\nc{\cR}{{\cal R}}
\nc{\cS}{{\cal S}}
\nc{\cX}{{\cal X}}
\nc{\cO}{{\cal O}}
\nc{\cU}{{\cal U}}

\begin{document}

\title{Einstein's Moon}

\author{D. Song}

\affiliation{School of Liberal Arts, Korea University of Technology \& Education, Chungnam 330-708, Korea}%


\begin{abstract}

An account of the subjective elements of quantum
mechanics or of whether, as Einstein famously asked, the
Moon exists when nobody is looking at it.

\end{abstract}


\maketitle

Einstein was not very happy with quantum theory, 
for a very good reason. Quantum theory is probabilistic 
at the fundamental level. Well, you may ask, what can be so 
wrong about the theory being probabilistic? Science is based on 
causality; that is, for every result, there is a cause. 
If the theory is probabilistic, it means the result appears to 
happen without a definite cause. This seems to be problematic as 
far as causality is concerned (see \cite{peres,QC} for a review). 

Moreover, this probabilistic nature of quantum theory happens 
when there is a measurement or observation. Again, what is the 
big deal with the theory involving the observation? Is science not 
all about experiments and observations? In case of quantum theory, 
the problem is that the observation often changes the status of the 
observed physical system. In other words, subjectivity is an essential 
element of quantum theory. This was something Einstein, and many other 
people, could not take.  They thought that science should provide a consistent 
truth about an objective reality rather than something that varies depending on a 
subjective perspective.  This sounds very reasonable. Or does it not?  

Experiments or observations form the basis of science. 
Although we often think science provides an objective law 
about physical systems, in fact, it yields a rule about the 
way we observe physical systems. This was true even before quantum 
theory, when distinction was not necessary to improve predictability; 
it only added extra burden. However, with the development of quantum 
theory at the beginning of the 20th century, the subjective aspect of 
science finally began to emerge and it started to matter, even in 
terms of actual predictability. In other words, quantum theory started to 
reach the ultimate limit of science, subjectivity.

The idea of subjectivity is nothing new. Philosophers have been talking 
about it for centuries. Descartes argued that at least the subjective 
thought itself was certain to exist, which he expressed in the well-known 
statement \lq\lq I think, therefore I am.\rq\rq Even in the 20th century, 
many philosophers discussed the subjective nature of existence itself. 
However, is it possible to argue this scientifically rather than 
philosophically? Is it possible to write down a precise and exact 
mathematical equation and show that existence is indeed subjective?   

In \cite{song}, it was shown that it is not possible to separate the observer 
from the observed using quantum theory. That is, physical systems, 
including atoms, the Moon, or the whole universe, do not exist in 
separate from my own existence. However, was the argument scientific? 
Was it mathematically precise and exact? The great power of quantum 
theory lies in its preciseness and exactness. That is, a state vector, 
a mathematical representation of the physical system, is a full and 
exact description. What comes next is even better. When observing 
the state vector, one needs to be in a certain reference frame, 
called an observable in quantum theory. An amazing part is that 
this reference frame is also full and exact, just like the state 
vector. Okay, one may argue, you can represent the physical system 
and the reference frame of the observer exactly, but this does not mean the universe is subjective.     

When you have this exact representation for the physical 
system and the observer, there is symmetry between the observer 
and the observed. Consider a rotational symmetry. That is, if the 
system were rotated clockwise or if you were rotated counterclockwise, 
you would observe exactly the same thing on both occasions.  
The symmetry between the object and the observer explains the 
phenomenon; it is called the Schr\"odinger and the Heisenberg picture 
in quantum theory. Why does this prove the universe is subjective? 

We experience some very strange phenomenon where this symmetry 
between the object and the observer breaks down. This phenomenon 
is consciousness!  In consciousness, one experiences the observation 
of one's own mental state, sometimes called self-awareness or reflexive 
self-consciousness. This is unique. The person is both the observer and 
the very object that is being observed. Because of consciousness, 
the symmetry, established on exact and precise mathematical 
representation of the object and the observer, is no longer valid. 
That is, one cannot separate the object from the observer. 
If the universe is the object that is being observed, then the universe has to be subjective as well.

Einstein once asked his young friend Abraham Pais if the Moon 
existed only when someone was looking at it \cite{pais}. Does the Moon, 
indeed, exist only when I observe it? If we assume that the Moon 
obeys quantum theory and the unique property of consciousness, 
as strange and counter-intuitive as it may seem, the Moon may 
not exist in separate from my own existence.

\end{document}